\documentclass[onecolumn, 11pt, letterpaper, showkeys, pra]{revtex4}
\usepackage{amstext}
\usepackage{amsmath}
\usepackage{amssymb}
\usepackage{graphicx}

\newcommand{\ud}{\mathrm{d}}
\newcommand{\vect}[1]{\boldsymbol{#1}}
\newcommand{\eq}[1]{Eq.~\eqref{#1}}
\newcommand{\fig}[1]{Fig.~\ref{#1}}
\newcommand{\stn}[1]{Sec.~\ref{#1}}
\newcommand{\be}{\begin{equation}}
\newcommand{\ee}{\end{equation}}
\newcommand{\ba}{\begin{align}}
\newcommand{\ea}{\end{align}}
\newcommand{\ti}[1]{\text{#1}}

\newcommand{\mc}[1]{\mathcal{#1}}

\begin{document}
\title{Single-molecule pulling: phenomenology and interpretation
 }
\author{Ignacio Franco}
\author{Mark A. Ratner}
\author{George C. Schatz}
\affiliation{Department of Chemistry, Northwestern University, Evanston, Illinois 60208-3113}

\begin{abstract}
Single-molecule pulling techniques have emerged as versatile tools for probing the noncovalent forces holding together the secondary and tertiary structure of macromolecules. They also constitute a way to study at the single-molecule level processes  that are familiar from our macroscopic thermodynamic experience. In this Chapter, we summarize the essential phenomenology that is typically observed during single-molecule pulling, provide a general statistical mechanical framework for the interpretation of the equilibrium force spectroscopy and illustrate how to simulate single-molecule pulling experiments using molecular dynamics. 

\textbf{Published in \emph{Nano and Cell Mechanics: Fundamentals and Frontiers},  edited by H.D. Espinosa and G. Bao (Wiley, Microsystem and Nanotechnology Series, 2013), Chap. 14, pages 359--388.}
\end{abstract}
\keywords{Molecular dynamics, macromolecules, mechanical properties, finite system thermodynamics, free energy reconstruction}
%\date{}
\maketitle
\tableofcontents
%\pagebreak

\section{Introduction}
\label{stn:intro}

In recent years a number of techniques have been developed that allow us to access the properties of single molecules (see, e.g., Refs.~\cite{ashkintweezers, rieftitin1997, Weissfluorescence, moerner99, Strick2000, ratnernitzan03,  ritort06, camden08}). These windows into the single-molecule world are teaching us with unprecedented detail how molecules behave, including how they move, their mechanical behavior, their conductivity, reactivity, optical properties, and other properties. 
While measurements made on bulk matter are averages over a whole ensemble of molecules, single-molecule  measurements highlight the contributions of the constituent parts to the ensemble.   A  key feature of the properties of single molecules is that fluctuations in the observables are comparable with the average values. This contrasts with macroscopic behavior (where such fluctuations are usually negligible) and indeed fluctuation-induced emergent phenomena can arise that, simply put, cannot be observed in the macroscale.

 In this Chapter we discuss basic aspects of single-molecule pulling experiments in which laser optical tweezers or atomic force microscopes (AFMs) are employed to exert mechanical forces on single molecules~\cite{rieftitin1997, Liphardt01, fernandez01, evans2001, bustamanterev05}, and the resulting stress/strain response is used as a probe of the molecular potential energy surface. The focus will be on how to simulate single-molecule pulling experiments using molecular dynamics, on what class of information can be extracted from the resulting force-extension isotherms and on how to interpret the basic phenomena typically observed upon pulling. 
 
 %--------------------------------------------------%
 \begin{figure}[htbp]
\centering
\includegraphics[width=0.3\textwidth]{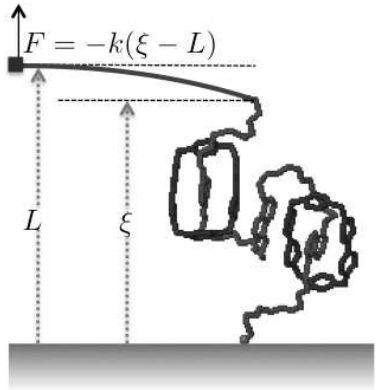} 
\caption{Schematic of a single-molecule pulling experiment using AFMs.  One end of the molecular system is attached to a surface and the other end to an AFM tip attached to a cantilever. During the pulling,  the distance between the surface and the cantilever $L$ is controlled while  the deflection of the cantilever from its equilibrium position measures the instantaneous applied force  $F(t) = -k (\xi(t) - L)$, where $k$ is the cantilever spring constant and  $\xi(t)$  the fluctuating molecular end-to-end extension.  }
     \label{fig:pullingfig}
\end{figure}

 %--------------------------------------------------%

 Figure~\ref{fig:pullingfig} shows a schematic of a single-molecule pulling experiment using AFMs (an equivalent setup exists for experiments that employ laser optical tweezers).  In it, one end of a molecule is attached to a surface, while the other end is attached to an AFM tip coupled to a cantilever. The distance between the surface and the cantilever $L$ is controlled while the molecular end-to-end distance $\xi(t)$ is allowed to fluctuate. The pulling is performed by varying $L(t)$ in some prescribed fashion. The force $F$ exerted during the process is determined by measuring the deflection of the cantilever from its equilibrium position $F= -k (\xi -L)$, where $k$ is the cantilever stiffness.   For a discussion of experimental aspects of single-molecule force spectroscopy  see Ref.~\cite{attila2008}.
 
The experiments exert mechanical control over the molecular conformation while simultaneously making thermodynamic measurements of any mechanically-induced unfolding events. 
 The stress maxima observed during pulling have been linked to the breaking~\cite{evans2001, freund} of noncovalent interactions, thus providing insight into the secondary and tertiary structures of macromolecules.   From the force-extension data it is possible to reconstruct the  molecular free energy profile along the extension coordinate, quantifying in this way thermodynamic changes undergone by the molecule during folding. 

The structure of this Chapter is as follows: Section \ref{stn:experiments} introduces the basic phenomenology observed during single molecule pulling through a discussion of representative experiments.  Section~\ref{stn:thermodynamics} establishes the statistical mechanical theory behind single-molecule pulling experiments and describes a method for reconstructing the molecular free energy profile along the extension coordinate from the force-extension data. Section~\ref{stn:modeling}, in turn, introduces direct and indirect methods that can be used to model single-molecule pulling experiments using molecular dynamics.  Section~\ref{stn:interpretation} focuses on the interpretation of single-molecule pulling experiments through an analysis of the basic structure of the molecular free energy profile.  Section~\ref{stn:conclusions} summarizes the main topics discussed in this Chapter.

\section{Force-extension behavior of single molecules}
\label{stn:experiments}

\begin{figure}[htbp]
\centering
\includegraphics[width=0.8\textwidth]{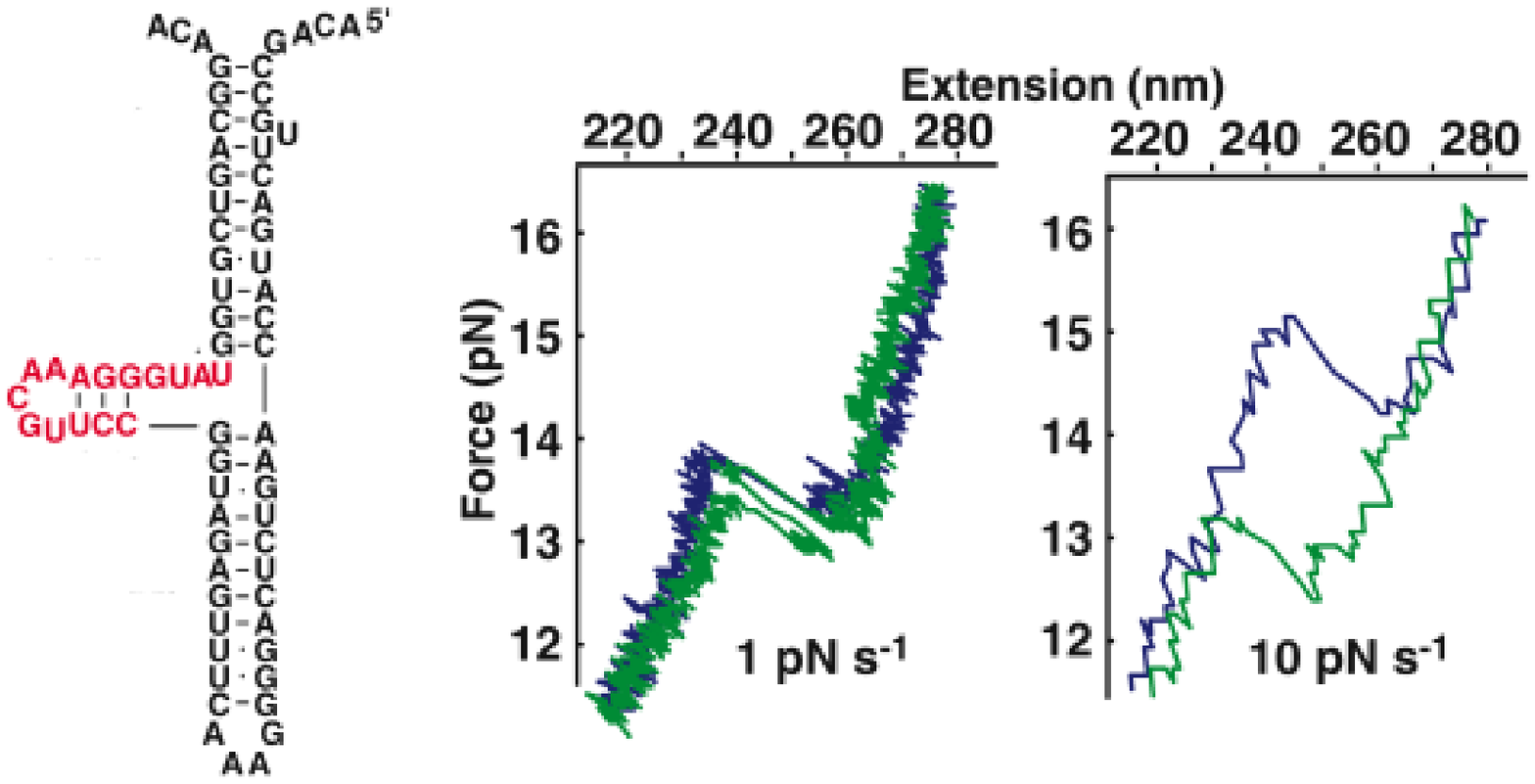} 
\caption{Force-extension behavior of a RNA hairpin obtained in an optical tweezer setup. The left panel shows the sequence and secondary structure of the main structural motif of the pulled RNA hairpin. The right plots detail the force-extension curves during stretching (blue) and contraction (green) when pulling under equilibrium (left) and nonequilibrium (right) conditions. Note that the $x$-axis is the molecular end-to-end distance $\xi$ and not the distance between the surface and the cantilever $L$. Figure adapted from~\cite{Liphardt01}. }
     \label{fig:rnapulling}
\end{figure}

In this section, we consider a series of examples of single-molecule pulling experiments that illustrate the basic phenomenology that can be observed.  Consider first the mechanically induced unfolding of a RNA hairpin~\cite{Liphardt01}. The hairpin is attached  to polystyrene beads via RNA/DNA hybrid handles and the pulling is performed in an optical tweezers arrangement. Figure~\ref{fig:rnapulling} shows the experimentally determined force-extension behavior and the details of  the hairpin structure. For fast pulling speeds hysteresis in the force-extension traces is evidenced by the fact that the traces observed upon pulling and subsequent contraction do not coincide. However, upon reduction of the pulling speed  the traces obtained during extension and  contraction do coincide, indicating that for all practical purposes the process occurs under reversible conditions. During pulling the force initially increases monotonically, then at $\sim 14$ pN it shows a drop, and then it increases again. In the process, the RNA hairpin undergoes a conformational transition from the hairpin structure to an extended coil. The drop in the force corresponds to the molecular unfolding event where the hydrogen bonding network holding the RNA hairpin together is broken.  The local maximum in the force-extension isotherm is interpreted as the force required to break the secondary structure of the molecule.

 %--------------------------------------------------%
%--------------------------------------------------%
 
The force-extension behavior of the RNA hairpin illustrates two basic phenomena that are commonly observed during single-molecule pulling. During pulling there is a region of \emph{mechanical instability} where the average force decreases with the extension $\frac{\partial \langle F \rangle_L}{\partial L} < 0$. This region occurs when the pulled molecule undergoes a conformational transition from a stable folded structure to an extended system. Note that around the unfolding region the molecule hops between the folded and the unfolded state. That is, around the region of mechanical instability the molecule plus cantilever system exhibits \emph{dynamical bistability} where the molecule unfolds and refolds, leading to blinking in the force measurement between a stronger force when the molecule is folded and a weaker force regime when extended. In \stn{stn:interpretation} we discuss the origin of these two phenomena.

 %--------------------------------------------------%
 \begin{figure}[htbp]
\centering
\includegraphics[width=1.0\textwidth]{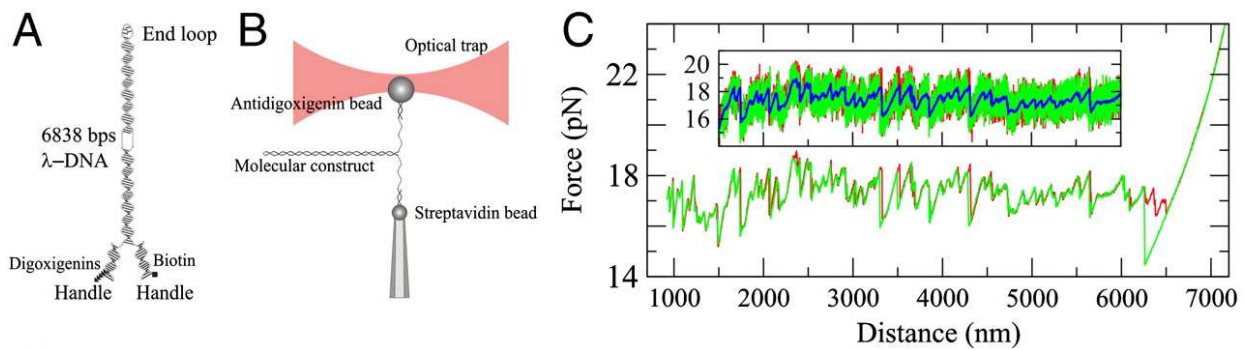} 
\caption{In this experiment the 6838 base pair long DNA hairpin schematically shown in (A) is pulled in the optical tweezer arrangement in (B). The pulling is done by moving the optical trap relative to the pipette at  low pulling speeds (10 nm/s). Panel (C) shows the force-extension profile during unfolding (in red) and refolding (in green). The inset shows the raw data obtained while the smooth curves are time-averaged. Except for some hysteresis in the last rip, the curves obtained during pulling and relaxing are almost identical. Figure adapted from~\cite{ritort2010}. }
     \label{fig:ritort}
\end{figure}
%--------------------------------------------------%
 
Figure \ref{fig:ritort}  shows the force-extension characteristics of a related system, a 6838 base-pair long DNA hairpin, also obtained in a optical tweezers arrangement~\cite{ritort2010}. Even for this complex system it is possible to find experimentally accessible pulling speeds (10 nm/s) where reversible behavior is recovered in the mechanically induced melting process. Melting in this case refers to dehybridization of the base pairs, and we see that unlike \fig{fig:rnapulling}, there is an extended unfolding region where the average force is in the 15-19 pN range while the length $L$ increases.  This corresponds to sequential dehybridization (believed to occur a few base pairs at a time) rather than a concerted breaking of all the base pairs at the same time.  The inset in the figure shows the raw data obtained in the experiment while the smooth curves shown are after further temporal averaging. The averaged data evidences the changes in the pulling force required to mechanically induce melting. It quantifies the strength of the type of hydrogen bonds encountered as the elongation proceeds.

%OJO need to ask George.

 The raw data evidences yet another fundamental aspect of this class of experiments:  the substantial thermal fluctuations in the force measurements that are typically observed during  pulling indicating that the system is not in the thermodynamic limit. 
 This observation has the important consequence of leading to a non-equivalence between different statistical ensembles.  In particular, it leads to a non-equivalence between the isometric ensemble (where  $L$ is controlled and  $F$ fluctuates) and the isotensional ensemble (where $F$ is controlled and $L$ fluctuates).  Throughout we will focus on the isometric ensemble since is the most common way to perform these experiments. For recent discussions of the nonequivalence between ensembles see~\cite{kreuzer2001, holm2009}.  See~\cite{kirmizialtin2005} for an interpretation of the pulling phenomenology in the isotensional ensemble.

 %--------------------------------------------------%
 \begin{figure}[htbp]
\centering
\includegraphics[width=0.5\textwidth]{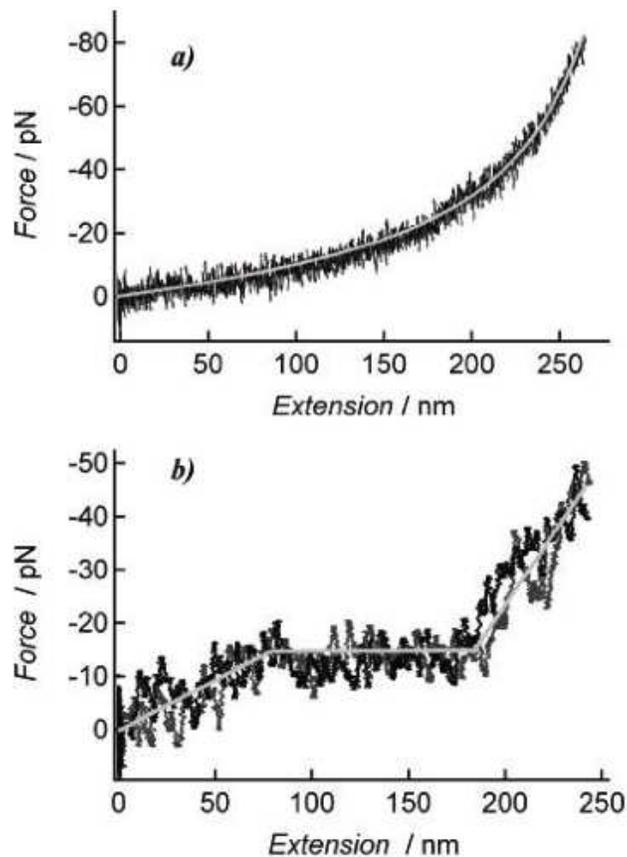} 
\caption{Reversible pulling of individual polystyrene chains in toluene (top panel) and water (bottom panel) obtained using an AFM setup.  Figure adapted from~\cite{walker07}. }
     \label{fig:walker}
\end{figure}
 %--------------------------------------------------%

As a third example, consider the reversible force-extension behavior of a synthetic polymer pulled in poor and good solvent conditions using an AFM~\cite{walker07}. Figure~\ref{fig:walker} shows the force-extension isotherm of an individual polystyrene (PS) chain pulled in toluene (good solvent) and water (poor solvent). In toluene (\fig{fig:walker}a),  the deformation of PS exhibits a worm-like chain~\cite{rubinsteinbook} behavior with a persistence length of (0.25 $\pm$ 0.05) nm.  In water (\fig{fig:walker}b) the behavior is qualitatively different. In this case, the force initially rises linearly up to $\sim 13$ pN,  then exhibits a plateau force at $\sim 13$ pN, followed by a third regime where the force rises again.  This behavior is well predicted by a model~\cite{halperin91} in which the initial linear restoring force is due to deformation of a collapsed spherical globule to an ellipsoid during which the polymer's surface energy increases. In this model, the plateau is associated with a conformational transition involving the coexistence of the globule and a stretched chain of polymer or polymer globules. For large stretching,  where coexistence is no longer possible, the force extension behavior of a Gaussian chain~\cite{rubinsteinbook} is recovered. Note that large-scale fluctuations in the force measurements are also evident in this system.

 %--------------------------------------------------%
\begin{figure}[htbp]
\centering
\includegraphics[width=0.5\textwidth]{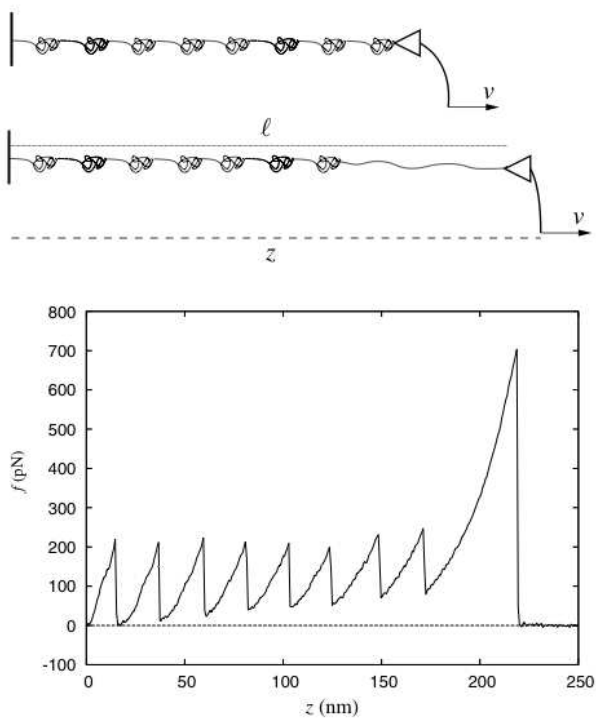} 
\caption{Top panel: schematic of the protein attached to the AFM tip. Bottom panel: force-extension behavior of the polyprotein composed of eight repeat of the Ig27 domain. Figure taken from~\cite{imparato08}.}
     \label{fig:imparato}
\end{figure}
 %--------------------------------------------------%

Consider last the forced unfolding of a recombinant polyprotein composed of eight repeats of the Ig27 domain of human titin obtained in an AFM setup~\cite{imparato08}. Figure~\ref{fig:imparato} shows the force-extension profile when pulling at a constant speed under nonequilibrium conditions. As the polyprotein is extended the force exerted on the molecule increases until one of the Ig27 domains suddenly unfolds. Unfolding of a domain reduces the overall force, but with increasing extension another local subunit unfolds. This occurs until all the domains are unfolded and the protein-tip interaction is broken. 
This results in the sawtooth pattern shown in \fig{fig:imparato} where each peak in the force-extension profile marks the unfolding of one subunit in the polyprotein and the last peak signals the detachment of the protein from the tip.

\section{Single-molecule thermodynamics}
\label{stn:thermodynamics}

In what follows we specify what we mean about thermodynamics in the context of single-molecule pulling experiments~\cite{bustamanterev05} and the class of thermodynamic information that is accessible from the force-extension isotherms. The thermodynamic system consists of the molecule plus cantilever coupled to a thermal bath, typically the surrounding solvent. As such, the equilibrium state of the system is well described by the canonical ensemble with configurational partition function
\be
\label{eq:partfuncL}
Z(L) = \int \ud \vect{r}\, \exp[-\beta U_L(\vect{r})],
\ee
where $\vect{r}$ denotes the coordinates of the $N$ atoms of the molecule, $\beta = 1/{k_\textrm{B}T}$ is the inverse temperature, and
\be
U_L(\vect{r}) = U_0(\vect{r}) + V_L[\xi(\vect{r})]
\ee
is the potential energy function of the molecule plus restraints. 
Here   $U_0(\vect{r})$  is the molecular potential energy plus potential restraints not varied during the experiment, and $V_L[\xi(\vect{r})]$ is the potential due to the cantilever at extension $L$. To a good approximation, the quantity $V_L[\xi(\vect{r})]$ is given by a harmonic function
\be
\label{eq:cantilever}
V_L[\xi(\vect{r})] = \frac{k}{2}\left[\xi(\vect{r}) - L\right]^2 
\ee
of stiffness $k$. The instantaneous force exerted by the cantilever on the molecule is determined by measuring the deflection of the cantilever from its equilibrium position and it is given by
\be
\label{eq:force}
F = - \nabla V_L =  \frac{\partial V_L}{\partial L} = - k\left[\xi(\vect{r}) - L\right],
\ee
where the gradient $\nabla$  is with respect to the $[\xi(\vect{r})- L]$ coordinate.
In writing Eqs.~\eqref{eq:cantilever} and~\eqref{eq:force}  the vector nature of  $L$, $\xi(\vect{r})$  and $F$ has been obviated since  these quantities are typically collinear in the single-molecule pulling setup.  For notational simplicity, in \eq{eq:partfuncL} and throughout this manuscript we ignore the constant multiplicative factor that makes partition functions dimensionless since they are irrelevant for determining  relative free energy changes. 

\subsection{Free energy profile of the molecule plus cantilever}

Because the state of the molecule plus cantilever under equilibrium conditions is well described by a canonical ensemble, the force exerted during the  pulling yields the associated change  in the Helmholtz free energy 
\be
\Delta A = A(L) -A(L_0)  = -\frac{1}{\beta}\ln \frac{Z(L)}{Z(L_0)}
\ee
 for the composite molecule plus cantilever system. To see this, note that in the canonical ensemble  the average force at a given extension $L$ is given by
\be
\label{eq:averageforce1}
\langle F \rangle_L =  \frac{1}{Z(L)} \int \ud \vect{r} \left(\frac{\partial V_L}{\partial L}\right) \exp\{-\beta U_L(\vect{r})\} = -\frac{1}{\beta} \frac{1}{Z(L)}\frac{\partial Z(L)}{\partial L} = \frac{\partial A}{\partial L}.
\ee
Therefore, if the pulling is done under reversible conditions, such that the state of the system is well described by a canonical ensemble at each step of the pulling,  the change in $A$ when pulling from $L_0$ to $L$ is determined by the reversible work exerted in the process
\be
\label{eq:helmholtz}
\Delta A   = \int_{L_0}^{L} \frac{\partial A}{\partial L'} \ud L'   =  \int_{L_0}^{L} \langle F \rangle_{L'}  \ud L'  = W_\text{rev}. 
\ee
Remarkably, even when the pulling is performed under nonequilibrium conditions it is still possible to determine the free energy changes during pulling by means of the Jarzynski non-equilibrium work fluctuation relation~\cite{jarzynski1, jarzynski2, jarzynski3}
\be
\label{eq:jarzynski}
\exp[-\beta\Delta A] = \langle \exp[-\beta W] \rangle_\textrm{non-eq} 
\ee
where the average in this expression is over nonequilibrium realizations of a given pulling protocol starting from a system initially prepared in the canonical ensemble.    \eq{eq:jarzynski} relates the nonequilibrium work with the equilibrium free energy changes of the system and holds arbitrarily far away from equilibrium.  Note that  by using Jensen's inequality ($e^{\langle x\rangle} \le \langle e^x \rangle$), \eq{eq:jarzynski} implies that the average nonequilibrium work is greater than or equal to the free energy change
\be
\langle W \rangle_\textrm{non-eq} \ge \Delta A,
\ee
which is the usual statement of the second law of thermodynamics

The above analysis associates a free energy change to the pulling process although  the system is not in the thermodynamic limit. This association relies on the assumption that the equilibrium state of the system is well described by a classical canonical ensemble.  This feature has been experimentally tested: In~\cite{Liphardt02} the authors experimentally demonstrated the validity of the Jarzynski equality  [\eq{eq:jarzynski}]. The Jarzynski equality supposes that the initial equilibrium state of the system is given by a classical canonical partition function and hence the test of the Jarzynski equality also tests whether this is a good description of the equilibrium state of the system. 

\subsection{Extracting the molecular potential of mean force $\phi(\xi)$}

The properties that are measured during the pulling are those of the molecule plus cantilever. Thus, direct integration of the force-extension isotherms yields the free energy profile of the composite system [see \eq{eq:helmholtz}]. However, what one is really interested in are the properties of the molecule itself. Specifically, one is interested in the molecular Helmholtz free energy profile (also known as the potential of mean force PMF~\cite{Kirkwood35}) along the extension coordinate $\xi$. The PMF succinctly captures thermodynamic changes undergone by the molecule during folding and determines the basic phenomenology that is observed upon pulling~\cite{kirmizialtin2005, franco09}.

\begin{figure}[htbp]
\centering
\includegraphics[width=0.8\textwidth]{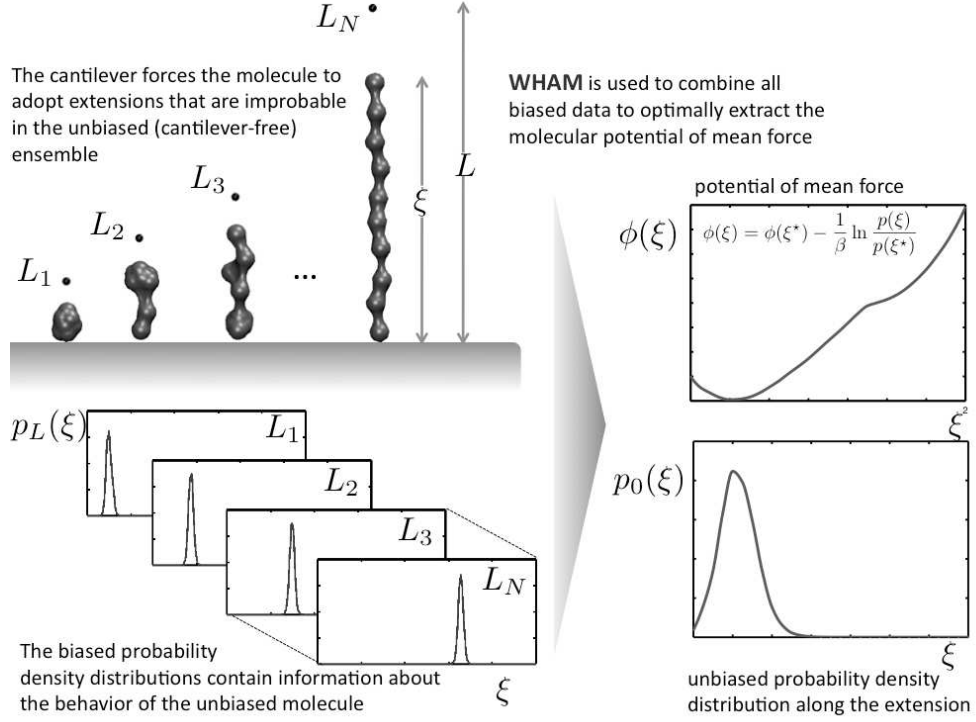} 
\caption{Scheme of the free energy reconstruction from single-molecule pulling. The potential of mean force $\phi(\xi)$  along the extension coordinate $\xi$ is determined by the probability density distribution $p_0(\xi)$ [\eq{eq:PMF}]. Sampling  $p_0(\xi)$ directly is challenging because there are regions along $\xi$ that have vanishingly small probability of being observed.  In a single-molecule pulling experiments the molecule is forced to adopt those improbable extensions $\xi$ by subjecting it to the bias due to the cantilever. For each $L$ visited ($L_1, L_2, \cdots, L_N$)  during pulling the experiments measure the probability density distribution along $\xi$ in the presence of the cantilever bias, $p_L(\xi)$. The $p_L(\xi)$ contain information about the unbiased probability density distribution $p_0(\xi)$ [\eq{eq:biasedvsunbiased}]. The WHAM combines all sampled $p_L(\xi)$ to properly remove the cantilever bias and extract $p_0(\xi)$ with minimal variance in the estimate.  }
     \label{fig:WHAM}
\end{figure}

A useful method to extract the molecular PMF  $\phi(\xi)$ from the force versus extension measurements is the Weighted Histogram Analysis Method (WHAM)~\cite{wham1,  wham2, whamroux, frenkelandsmit, mcwham}.  The WHAM analysis combines all the force-extension data collected during the pulling to properly remove the bias due to the cantilever and  extract the PMF.  In fact, one can view the single-molecule pulling process as an experimental realization of the WHAM methodology using harmonic potential restraints. We now describe how to extract the PMF from equilibrium force-extension data using the WHAM.  A schematic of the process is shown in \fig{fig:WHAM}. 

 %--------------------------------------------------%
 %--------------------------------------------------%

The PMF is defined by~\cite{whamroux}
\be
\label{eq:PMF}
\phi(\xi)  = \phi(\xi^\star) -\frac{1}{\beta}\ln \left[ \frac{p_0(\xi)}{p_0(\xi^\star)} \right],
\ee
 where $\xi^\star$ and  $\phi(\xi^\star)$ are arbitrary constants and 
\be
\label{eq:unbiased}
 p_0(\xi) = \frac{\int \ud \vect{r}\,  \delta[\xi-\xi(\vect{r})] \exp[-\beta U_0(\vect{r})]}{ \int \ud \vect{r} \,\exp[-\beta  U_0(\vect{r}) ] }  = \frac{Z_0(\xi)}{Z_0} = \langle\delta[\xi- \xi(\vect{r})]\rangle.
\ee
Here, 
\be
Z_0 = \int \ud \vect{r} \,\exp[-\beta  U_0(\vect{r}) ]
\ee
 is the  configurational partition function of the molecule, and
\be
\label{eq:partdens}
Z_0(\xi) = \int \ud \vect{r}\,  \delta[\xi-\xi(\vect{r})] \exp[-\beta U_0(\vect{r})].
\ee
In the context of the pulling experiments  $p_0(\xi)$ is the probability density that the molecular end-to-end distance function  $\xi(\vect{r})$  adopts the value $\xi$ in the unbiased (cantilever-free) ensemble.  The quantity   $p_0(\xi)$ determines the PMF  up to a  constant and can be estimated from the force measurements, as we now describe.

At this point it is  convenient to discretize the pulling process and suppose that $M$ extensions $L_1, \ldots, L_i, \ldots, L_M$ are visited during pulling. The potential  of the system plus cantilever at extension  $L_i$ is   $U_i = U_0 + V_i$, where $V_i = \frac{k}{2}[\xi(\vect{r}) - L_i]^2$ is the potential bias due to the cantilever. In the $i$-th biased measurement knowledge of the force $F_i$ and length $L_i$ gives the molecular end-to-end distance.  The probability density of observing the value  $\xi$  at this given extension is given by
\be
\label{eq:biased}
p_i(\xi) = \frac{\int \ud \vect{r}\, \delta[\xi-\xi(\vect{r})] \exp[-\beta U_i(\vect{r})]    }{\int \ud \vect{r}  \, \exp[-\beta U_i(\vect{r})]  } =  \frac{Z_i(\xi)}{Z_i} = \langle \delta[ \xi - \xi(\vect{r})] \rangle_i,
\ee
where  $Z_i= \int \ud \vect{r}\,  \exp[-\beta U_i(\vect{r})]$ is the configurational partition function for the system plus cantilever at the $i$-th extension, and
\be
Z_i(\xi) = \int \ud \vect{r} \delta[\xi-\xi(\vect{r})] \exp[-\beta U_i(\vect{r})].
\ee
In principle, the unbiased probability density $p_0(\xi)$ can be reconstructed from each $p_i(\xi)$  since, by virtue of Eqs.~\eqref{eq:unbiased} and~\eqref{eq:biased}, these two quantities are  related by 
\be
\label{eq:biasedvsunbiased}
p_0(\xi)  =  \exp[ + \beta V_i(\xi)] \frac{Z_i}{Z_0} p_i(\xi),
\ee
where we have exploited the fact that  $Z_0(\xi)= \exp[+\beta V_i(\xi)]Z_i(\xi)$.  In practice this approach will not work because the range of $\xi$ values where $p_0(\xi)$ and $p_i(\xi)$ differ significantly from zero need to overlap~\cite{frenkelandsmit}. Nevertheless, the experiments  do provide  measurements at a wealth of  values of $L_i$ that can be combined  to estimate  $p_0(\xi)$:
\be 
p_0(\xi) = \sum_{i=1}^{M} w_i \exp[+\beta V_i(\xi)]\frac{Z_i}{Z_0} p_i(\xi),
\ee
where the $w_i$ are some normalized ($\sum_i w_i = 1$) set of weights. In the limit of perfect sampling any set of weights should yield  the same $p_0(\xi)$. In practice, for finite sampling it is convenient to employ a set that minimizes the  variance in the  $p_0(\xi)$ estimate  from the series of independent estimates of biased distributions. Such a set of weights  is precisely provided by the WHAM prescription~\cite{wham1, wham2, frenkelandsmit}
\be
\label{eq:weights}
w_i  = \frac{\exp[-\beta V_i(\xi)] \dfrac{Z_0}{Z_i g_i}}{\sum_{i=1}^{M}  \exp[-\beta V_i(\xi)] \dfrac{Z_0}{Z_i g_i}},
\ee
where $g_i = 1+2\tau_i$ is the statistical inefficiency~\cite{wham3}  and  $\tau_i$   the integrated autocorrelation time~\cite{wham1, wham3}. The difficulty in estimating the $\tau_i$ for each measurement generally leads to further supposing that the $g_i$'s are approximately constant  and  factor out of \eq{eq:weights}, so that
\be
\label{eq:wham}
p_0(\xi)  = 
\frac{\sum_{i=1}^{M} p_i(\xi)}{\sum_{i=1}^{M}  \exp[-\beta V_i(\xi)] Z_0/Z_i}.
\ee
Neglecting the $g_i$'s from \eq{eq:wham} does  not imply that the resulting estimate of  $p_0(\xi)$ is incorrect; but simply that  the weights selected do not precisely minimize the variance in the estimate. Experience with this method indicates that  if the $g_i$'s do not differ by more than an order of magnitude their effect on $\phi(\xi)$ is small~\cite{wham2, gwham}.

The computation of the PMF then proceeds as follows. Suppose that $N_i$ force measurements are done for each extension  $i$.   Knowledge of the force $F_i^j$ ($j=1, \ldots, N_i$) and of $L_i$ gives the molecular end-to-end distance $\xi_i^j$ in each of these measurements.  The numerator in \eq{eq:wham} is then estimated by  constructing a histogram with all the available data,
\be
\sum_{i=1}^{M} p_i(\xi) \approx \sum_{i=1}^{M}\sum_{j=1}^{N_i} \frac{C_i^{j}(\xi)}{N_i \Delta\xi},
\ee
where $\Delta\xi$  is the bin size,  and $C_i^j(\xi)= 1$ if $\xi_i^j\in [\xi - \Delta\xi/2, \xi + \Delta\xi/2)$ and zero otherwise.  Estimating the  denominator in \eq{eq:wham} requires knowledge of the $Z_i$'s, the configurational partition functions of the system plus cantilever at all extensions $\{L_i\}$ considered. There are two ways to obtain these quantities. If reversible force-extension data is available, the most direct one is to employ the Helmholtz free energies for the system plus cantilever obtained through the thermodynamic integration in \eq{eq:helmholtz}, as they determine  the $Z_i$'s  up to a constant multiplicative factor. Alternatively,  it is also possible to determine the ratio of the configurational partition functions between  the $i$-th biased system and its  unbiased counterpart using $p_0(\xi)$:
\be
\label{eq:partitionf}
\frac{Z_i}{Z_0}  = \int \ud\xi\, \exp[-\beta V_i(\xi)] p_0(\xi).
\ee
Equations~\eqref{eq:wham} and~\eqref{eq:partitionf}  can be solved iteratively. Starting from a guess for the $Z_i/Z_0$, $p_0(\xi)$ is estimated using \eq{eq:wham} and normalized. The resulting $p_0(\xi)$ is then used to obtain a new set of $Z_i/Z_0$ through \eq{eq:partitionf}, and the process is repeated until self-consistency. This latter approach is the usual procedure to solve the WHAM equations.  For computational implementations of the WHAM methodology see Refs.~\cite{molpull, grossfieldwham, minhferbe, gwham}.

\subsection{Estimating force-extension behavior from $\phi(\xi)$}
\label{stn:forcefrompmf}

The PMF is central to the interpretation of the phenomena observed upon pulling and in the estimation of the force-extension behavior when pulling with cantilevers of arbitrary stiffness~\cite{franco09}. Its utility relies on the fact that the  configurational partition function of the molecule plus cantilever at extension $L$ [\eq{eq:partfuncL}]  can be expressed as
\be
\label{eq:partfuncL2}
Z(L) = \int\ud {\xi}\, \exp\{-\beta[\phi(\xi) + V_L(\xi)] \}, 
\ee
where we have neglected the  constant and $L$-independent multiplicative factor that arises in the transformation since it is irrelevant for the present purposes. The above relation implies that the extension process can be viewed as thermal motion along  a one-dimensional effective potential determined by the PMF $\phi(\xi)$ and the bias due to the cantilever $V_L(\xi)$:
\be
\label{eq:effectivepotential}
U_L(\xi) = \phi(\xi) + V_L(\xi).
\ee
Further, by means of the PMF  it is  possible to estimate the force vs. extension characteristics for the composite system for  any value of the force constant $k$. This is because  the average force exerted on the system at extension $L$ can be expressed as:
\be
\label{eq:averageforce}
\langle F \rangle_L = - \frac{1}{\beta} \frac{\partial  \ln[Z(L)/Z_0]}{\partial L} 
  = \frac{\int \ud\xi \, \frac{\partial V_L(\xi)}{\partial L}\exp\{ -\beta [\phi(\xi) + V_L(\xi)]\}   }
{ \int \ud\xi\, \exp \{ -\beta [\phi(\xi) + V_L(\xi)]\}   },
\ee
 where, for convenience, we have  introduced $Z_0$ in the logarithm just to make the argument  dimensionless. Since $\phi(\xi)$   is  a property of the isolated molecule it is  independent of $k$. Hence,   \eq{eq:averageforce} can be employed to estimate the $F$-$L$ isotherms for arbitrary $k$.  Another quantity of interest is the probability density in the force measurements as a function of the extension $L$, given by
 \be
 \label{eq:Fprobdens}
 p_L(F) =  \langle \delta[ F - F(\xi)] \rangle_L =  \frac{\int \ud\xi \,\delta[F - F(\xi)] 
 \exp\{ -\beta [\phi(\xi) + V_L(\xi)]\}   }
{ \int \ud\xi\, \exp \{ -\beta [\phi(\xi) + V_L(\xi)]\}   },
\ee
as this quantity highlights the fluctuations in the force measurements during pulling.

\section{Modeling single-molecule pulling using molecular dynamics}
\label{stn:modeling}

\subsection{Basic computational setup}

 %--------------------------------------------------%
 \begin{figure}[htbp]
\centering
\includegraphics[width=0.3\textwidth]{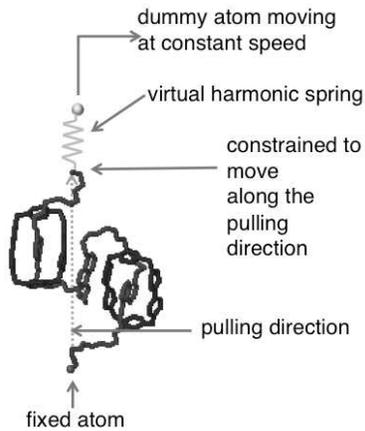} 
\caption{Setup of a MD pulling simulation. In it, one end of the molecule is fixed while the opposite end is attached to a dummy atom via a virtual harmonic spring.  The position of the dummy atom is the simulation analogue of the cantilever position and is controlled throughout. The pulling is performed by moving the dummy atom away from the molecule at a constant speed. The fluctuating deflection of the harmonic spring measures the force exerted during the pulling and is the simulation analog of the cantilever deflection in the AFM pulling experiment.}
     \label{fig:pullingscheme}
\end{figure}
 %--------------------------------------------------%

Computationally, the forced unfolding of single molecules can be studied using ``pulling'' or ``steered" molecular dynamics (MD) simulations~\cite{smd1, smd2, grubmuller99}. The general setup of this type of simulations is schematically described in \fig{fig:pullingscheme}. The stretching computation begins by  attaching one end of the molecule to a rigid isotropic harmonic potential that mimics the molecular attachment to the surface. Simultaneously, the opposite molecular  end is connected  to a dummy atom via a virtual harmonic spring.  The position of the dummy atom is the simulation analogue of  the  cantilever position $L$, and is controlled throughout. In turn,   the  varying deflection of the virtual harmonic spring   measures the force exerted during the pulling.  The stretching is caused by moving the dummy atom away from the molecule at a constant speed.  The pulling direction is defined by  the  vector  connecting the two terminal atoms  of the complex.   Since cantilever potentials are typically stiff in the direction perpendicular to the pulling, in the simulations the terminal atom that is being pulled is typically forced  to move along the pulling direction by introducing appropriate additional restraining potentials. 

In the simulations, the molecule plus any surrounding solvent are assumed to be weakly coupled to a heat bath and the effect of the heat bath is modeled through a thermostat. Thermostats that are useful in simulating the force spectroscopy include  the Nos\'e-Hoover chain~\cite{nh1, nh2, nh3, frenkelandsmit} and the Langevin thermostat~\cite{thijssenbook}. In turn, velocity rescaling thermostats, like the Berendsen thermostat~\cite{berendsen}, can be problematic because they can lead to a spurious violation of the second law  during  the pulling because the ensemble generated by them does not  satisfy the equipartition theorem~\cite{flyingicecube}.  The thermal dynamics of the molecule, cantilever and any surrounding solvent are then  followed in a NVT or NPT ensemble using MD.  Freely available MD codes such as NAMD~\cite{namd} and GROMACS~\cite{gromacs} have pulling routines implemented in them. We have, in addition, developed~\cite{franco09} a pulling routine for TINKER~\cite{tinker}  and  created a  graphical interface for it called MOLpull~\cite{molpull}. MOLpull   is accessible through the NanoHub (http://nanohub.org/) and can be used to perform the pulling and reconstruct the PMF along the extension coordinate of arbitrary molecules.

\subsection{Modeling strategies}

One of the challenges in simulating pulling experiments  using MD is to bridge the  large  disparity between the pulling speeds $v$ that are  employed  experimentally and those that can be accessed computationally.   While typical experiments often use   $v \sim 1-10^{-3}$ $\mu$m/s, current computational capabilities require pulling speeds that  are several ($\sim$5 to 8) orders  of magnitude faster.  

One possible strategy  that can be used to overcome this difficulty is to strive for  pulling speeds that are slow enough that   reversible behavior is recovered. Under such conditions   the results become independent of $v$, and the simulations comparable to experimental findings.  The way to determine if the pulling is done under quasistatic conditions is by making sure that the force-extension curves during extension and subsequent contraction essentially coincide, such that the net work exerted during a pulling/retraction cycle is negligible. The advantage of this approach is that there is a systematic way to test if quasistatic behavior is recovered by estimating the amount of work dissipated  in a pulling/retraction cycle.  Its limitation is that the pulling speeds that are required to obtain reversible behavior may be computationally unfeasible for pulling large macromolecules.  In fact, the examples that exist in the literature where reversible behavior has been computationally recovered are for relatively small molecular systems~\cite{franco09, stacker, azogrubmuller} using pulling speeds between $10^{-4}-10^{-7}$ nm/ps.  When pulling large macromolecules, present computational limitations require using pulling speeds of $\sim10^{-3}$ nm/ps~\cite{schulten10} which is far too fast for recovering reversible behavior even in simple systems.

A second possible strategy is to pull  under irreversible conditions. A single nonequilibrium trajectory can recover some qualitative structural features encountered during the pulling~\cite{schulten10, grubmuller99}, but it is not sufficient to reconstruct the molecular potential of mean force. In order to reconstruct the PMF it is necessary to pull many times and use a nonequilibrium work fluctuation theorem~\cite{jarzynski1,  crooks99, jarzynski2, jarzynski3} to reconstruct the free energy profile~\cite{hummer01,  park04,  hummer05,  kiang,    imparato08, minh08, pohorille2010, Hummer10}. The advantage of this strategy is that it can be used when equilibrium data is difficult to obtain, for example during the forced unfolding of a complex protein like the one in \fig{fig:imparato}. Note, however, that this strategy requires adequately estimating the average of an exponential quantity (\eq{eq:jarzynski}) that often has poor convergence properties~\cite{jarzynskiconvergence, pohorille2010}. For a computational implementation  of some nonequilibrium methods to reconstruct the PMF see~\cite{minhferbe}.

As a third \emph{indirect} alternative it is possible to reconstruct the molecular potential of mean force without pulling continuously but rather by sampling the fluctuating force at selected points along the extension coordinate.  For each $L$ the system is allowed to thermally equilibrate and the force is measured for a given amount of simulation time. The data obtained is used to reconstruct the PMF using WHAM  as described in \stn{stn:thermodynamics}, and then the PMF is employed to estimate the force-extension behavior of the molecule plus cantilever. 
 While the free energy reconstruction using continuous equilibrium force-extension data uses a data set with a large (essentially continuous) set of $L$ values with very few data points for the force at each $L$, this strategy works in a different limit where one has few extensions $L$ but extensive sampling of the force at each extension. The advantage of this procedure is that it can be used to reconstruct the PMF of molecules that are computationally challenging to pull under equilibrium or near equilibrium conditions.  The limitation is that there is no entirely satisfactory way to judge when the system has achieved a state of thermal equilibrium in the MD trajectory. This contrasts with the direct strategy involving explicit pulling in which it is easy to judge when quasistatic behavior has been recovered by measuring the amount of dissipative work in extension/contraction cycles. As in the direct equilibrium pulling, the sampling for each $L$ has to be sufficient to capture all relevant molecular events that are consistent with the extension.

Another important aspect that has to be taken into consideration when modeling single-molecule pulling is the cantilever stiffness employed.   Experimentally accessible AFM cantilevers stiffnesses are in the $10^{-1}-10^{4}$ pN/\AA~ range~\cite{attila2008}. While experiments tend to favor soft cantilever potentials because they provide high resolution in the force measurements,  in the simulations stiff cantilevers are preferable. This is because when employing stiffer cantilever potentials: (i) a shorter range of $L$ values is required to explore all the molecular extension coordinates thus reducing the simulation time; (ii) the possible values of $\xi$ that are accessible for the molecule for a given $L$ are further restricted making it faster to sample all accessible conformational space.  As a consequence of (ii), the amount of dissipative work in a pulling/retraction cycle decreases with increasing cantilever stiffness for fixed pulling speed~\cite{marsili}.   In our experience with molecular pulling, a rule of thumb that we frequently use is that a cantilever stiffness of $\sim~1$ pN/\AA~ can be regarded as soft and one with stiffness of $\sim10^{2}$ pN/\AA~ as stiff.  These numbers, however, will clearly depend on the particular system that is being pulled.

\subsection{Examples}

\subsubsection{Direct modeling: Oligorotaxanes}
\label{stn:oligorotaxane}

\begin{figure}[htbp]
\centering
\includegraphics[width=0.7\textwidth]{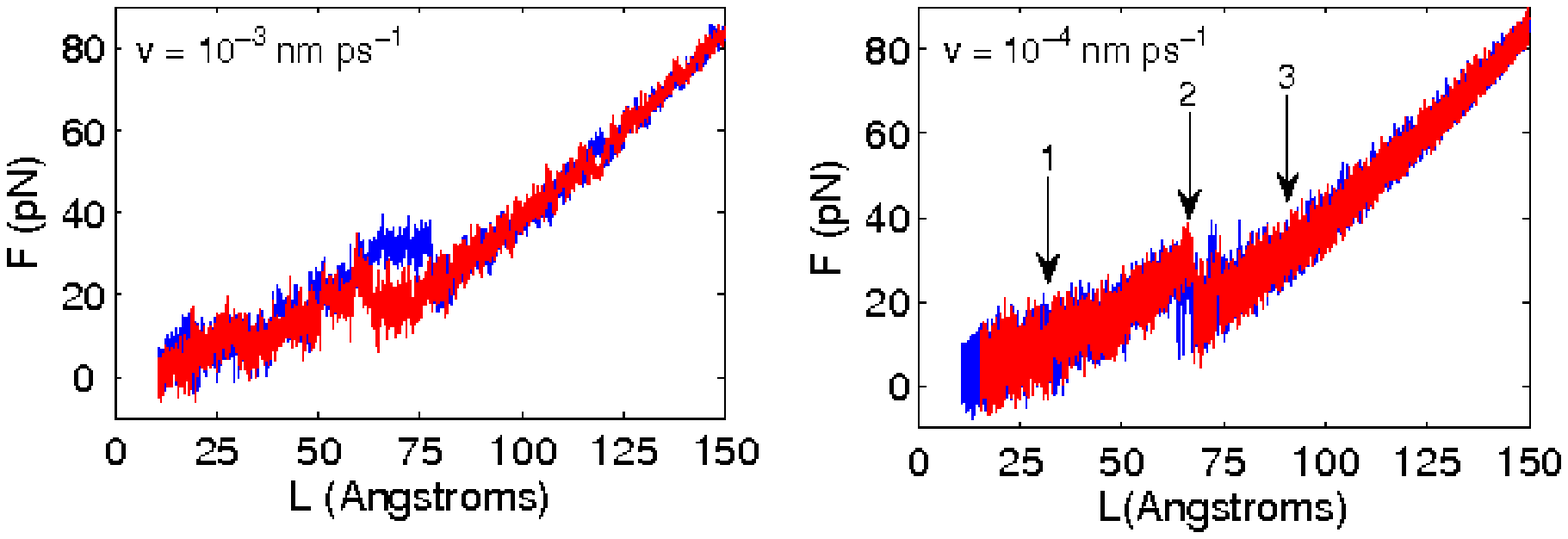} 
1\includegraphics[scale=0.3]{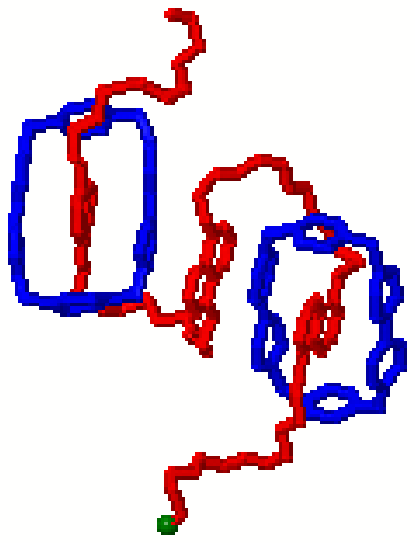}
2\includegraphics[scale=0.3]{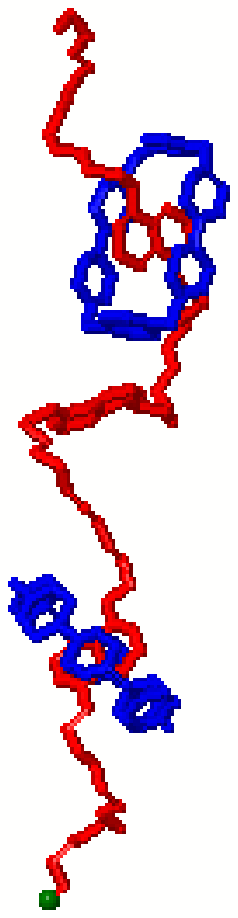}
3\includegraphics[scale=0.3]{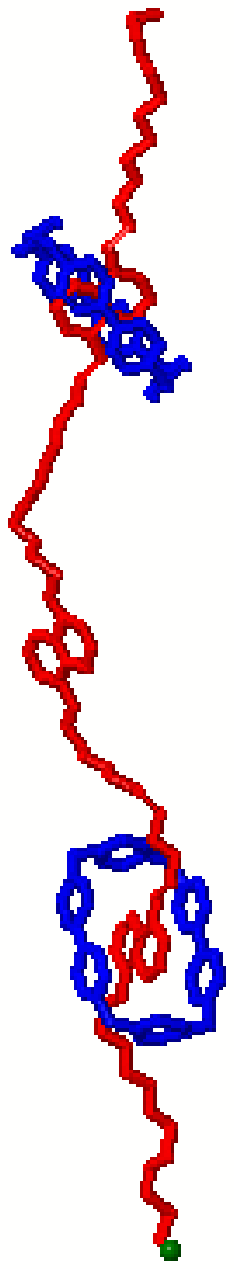}
\caption{ Simulated force-extension behavior of a [3]rotaxane  immersed in a high dielectric medium at 300K obtained using MD and the MM3~\cite{MM31} force field. The plots show the force-extension curves during stretching (blue) and retraction (red) using a pulling speed $v$ of $10^{-3}$ and $10^{-4}$ nm/ps. The  cantilever employed has a soft spring constant of  $k=1.1$ pN/\AA.  Snapshots of structures (labels 1, 2 and 3) encountered during pulling  are shown in the far right. }
     \label{fig:approachtoeql}
\end{figure}

As an example of an explicit simulation of a single-molecule pulling experiment consider the forced unfolding of a [3]rotaxane~\cite{franco09, francoenergetics, Basu11}.   Figure~\ref{fig:approachtoeql} shows the force-extension curves during the extension (blue)  and subsequent contraction (red) for two different pulling speeds and a schematic of the structure of the molecule that is being pulled.  The oligorotaxane\footnote{To be precise, we actually deal with pseudorotaxanes that lack
the steric blocking units at the chain termini that force the
cyclophanes to remain attached to the oligomeric chain. For simplicity,
we will use the term rotaxane.} consists of two cyclobis(paraquat-\emph{p}-phenylene) tetracationic cyclophanes (the blue boxes)  threaded onto a linear chain (in red) composed of three naphthalene units linked by oligoethers with oligoether caps at each end.  As the system is stretched the oligorotaxane undergoes a conformational transition from a folded globular state to an extended coil. In the process, the force initially  increases approximately linearly with $L$,  then  drops, and subsequently increases again.   For the faster pulling speed hysteresis in the force-extension cycle is evident while for  $v=10^{-4}$ nm/ps  the  $F$-$L$ curves obtained during extension and contraction essentially coincide and, for all practical purposes, the pulling occurs under equilibrium conditions. Note that thermal fluctuations in the force measurements are roughly comparable to the average values indicating that the system is far from the thermodynamic limit.  

\begin{figure}[htbp]
\centering
\includegraphics[width=0.7\textwidth]{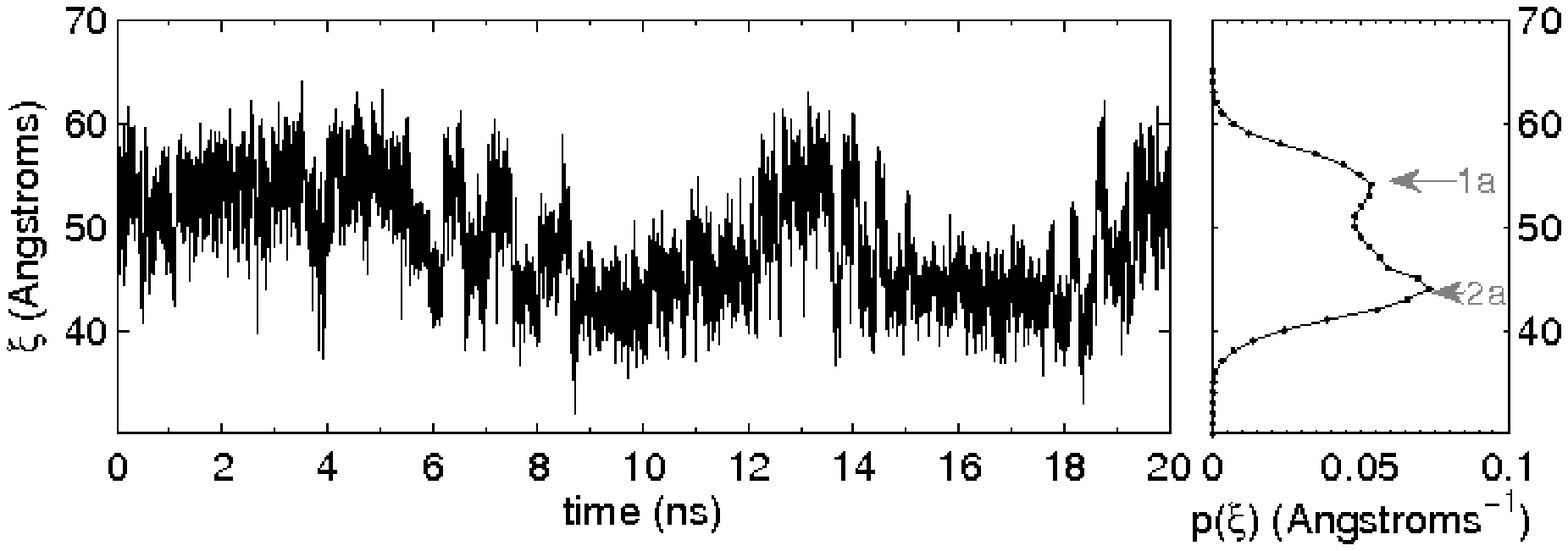} 
1a\includegraphics[scale=0.3]{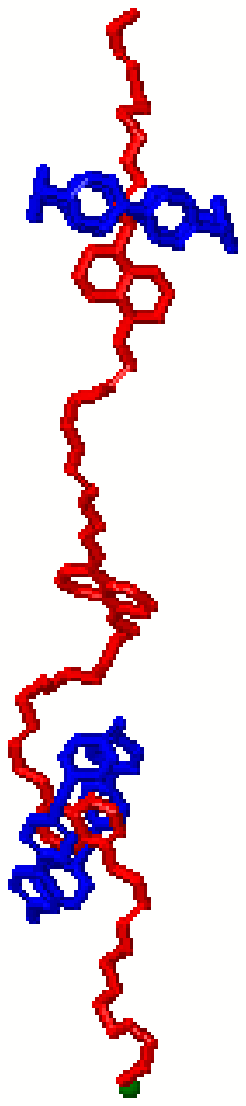}
2a\includegraphics[scale=0.3]{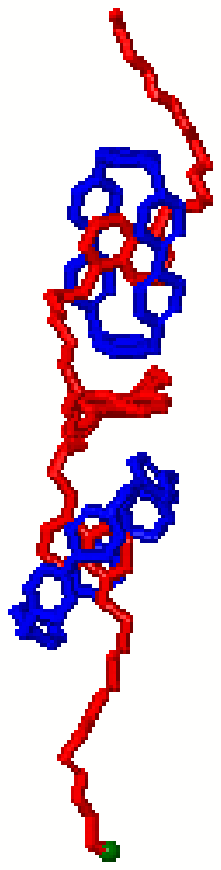}
\caption{Dynamical bistability when $L$ is fixed around the region of mechanical instability. The figure shows the time dependence and probability density distribution  of the end-to-end molecular extension $\xi$ when the [3]rotaxane plus cantilever is constrained to reside in the unstable region of \fig{fig:approachtoeql}, with $L=70.0$ \AA. Typical structures encountered in this regime (labels 1a-2a) are  shown in the right.}     \label{fig:3DNP2ring.unstable}
\end{figure}

The force-extension behavior predicted by the MD simulations for the oligorotaxane is qualitatively similar to that experimentally observed for the RNA hairpin (recall \fig{fig:rnapulling}). Specifically, there is a region of mechanical instability where the force drops  with increasing extension ($\frac{\partial \langle F \rangle_L}{\partial L} < 0$). Further, if one fixes $L$ around the region of mechanical instability ($L=70.0$~\AA~ in this case) and allows the system to evolve, the molecule undergoes conformational transitions between  a  folded globular state and  an extended coil, see \fig{fig:3DNP2ring.unstable}.  The right panel in  \fig{fig:3DNP2ring.unstable} shows the probability density of the distribution of  $\xi$ values obtained from a 20 ns trajectory. The system exhibits a clear dynamical bistability along the $\xi$ coordinate. As in the RNA case, the bistability along the end-to-end distance  leads to  blinking in the force measurements from a high force to a low force regime during the pulling.

%----------------------------------------------------------------%
%----------------------------------------------------------------%

%----------------------------------------------------------------%
%----------------------------------------------------------------%

 %--------------------------------------------------%
\begin{figure}[htbp]
\centering
\includegraphics[width=0.4\textwidth]{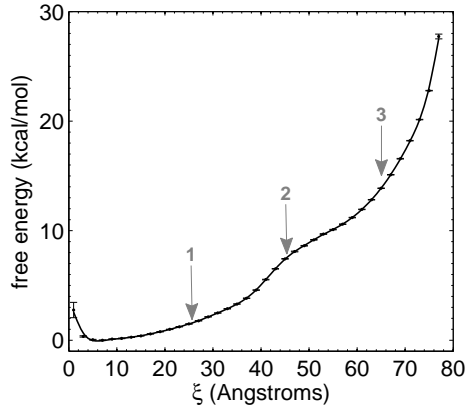} 
\caption{Molecular potential of mean force for the [3]rotaxane along the end-to-end distance $\xi$.  The solid lines result  from a spline interpolation of the available data points. Typical structures (labels 1-3) are shown in  
\fig{fig:approachtoeql}. }
     \label{fig:rotaxanePMF}
\end{figure}
 %--------------------------------------------------%

The molecular PMF of the [3]rotaxane reconstructed from the force-extension data is shown in \fig{fig:rotaxanePMF}. The thermodynamic native state of the oligorotaxane, i.e. the minimum in $\phi(\xi)$, corresponds to a  globular folded structure with an average end-to-end distance of 7~\AA. The PMF consists of a convex region (a region of positive curvature) for $\xi< 44$ \AA~ when the molecule is folded, another convex region for $\xi>52$ \AA~when the molecule is extended and a region of concavity  (where $\frac{\partial^2\phi}{\partial \xi^2} < 0$) for $\xi= 44-52$ \AA~ where the conformational transition occurs. The region of concavity along the PMF arises because the molecule is inherently bistable along some natural unfolding pathway. 
As discussed in \stn{stn:interpretation}, this characteristic structure of the PMF is responsible for many of the interesting features observed during pulling.

\subsubsection{Indirect modeling: DNA dimer}

 %--------------------------------------------------%
\begin{figure}[htbp] 
\centering
\includegraphics[width =0.4\textwidth]{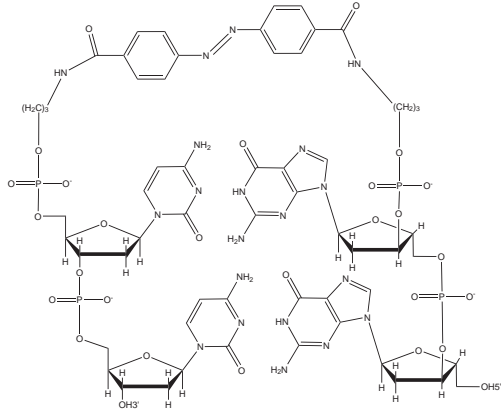} 
   \caption{DNA dimer consisting of two guanine-cytosine base pairs connected by a \emph{trans}-azobenzene linker.  }  
   \label{fig:dnaschematic} 
   \end{figure}
 %--------------------------------------------------%

As an example of indirect modeling of the force spectroscopy, we now consider the forced extension of the DNA dimer shown in \fig{fig:dnaschematic} along the O5'-O3' coordinate in explicit water~\cite{dnamotor}.  The dimer is composed of two guanine-cytosine base pairs capped by a  \emph{trans}-azobenzene linker.  
Computationally, it is  described by the CHARMM27 force field and the total system is composed of the 178 DNA hairpin atoms, 4 sodium ions, and 1731 TIP3P water molecules in a $27.1$\AA$\times34.7$\AA $\times56.3$\AA~box in the NVT ensemble at 300K (see Ref.~\cite{dnamotor} for details).
Even for this relatively simple system it is challenging to computationally obtain equilibrium force-extension data by pulling directly because the pulling has to be slower than any characteristic timescale of the system. An indirect approach is thus  more convenient.

The sampling along the extension coordinate required to reconstruct the potential of mean force proceeded as follows:   The distance, $L$, between the surface and the cantilever was fixed at several different values along the extension coordinate.  The system was first allowed to equilibrate for 4 ns. Subsequently, the dynamics was followed for 8 ns and the end-to-end distance $\xi$ recorded every 1 ps.  The cantilever was taken to have a stiffness of $k_0=110$ pN/\AA.
In total, 155 extensions were simulated.  Values of $L$ ranged from $6.75$ \AA ~to $44.75$ \AA ~while $\xi$ ranged from $2.9$ \AA ~to $44.9$ \AA.  Total analyzed simulation time was $1.24$ $\mu$s.

 %--------------------------------------------------%
\begin{figure}[htbp] 
\centering
\includegraphics[width =0.8\textwidth]{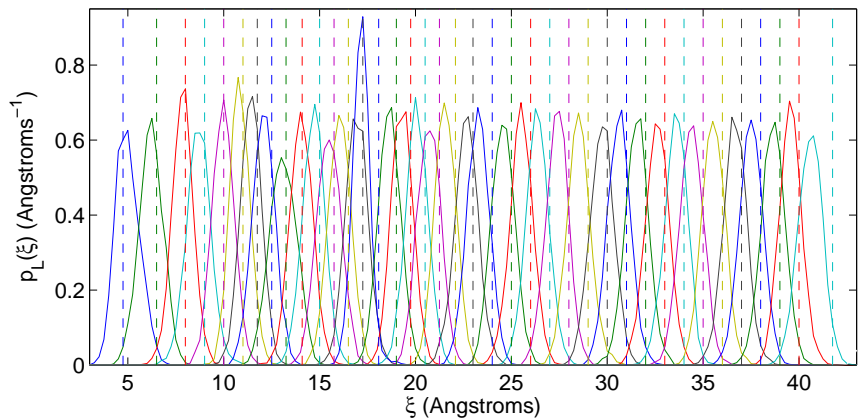} 
   \caption{Probability density distribution of $\xi$ when the cantilever bias is fixed at position $L$ for the DNA dimer. Each curve corresponds to a different $L$. The dotted lines mark the $L$ chosen in each case. }  
   \label{fig:histogram} 
   \end{figure}
 %--------------------------------------------------%
 
\begin{figure}[htbp]
   \centering
   \includegraphics[width =0.65\textwidth]{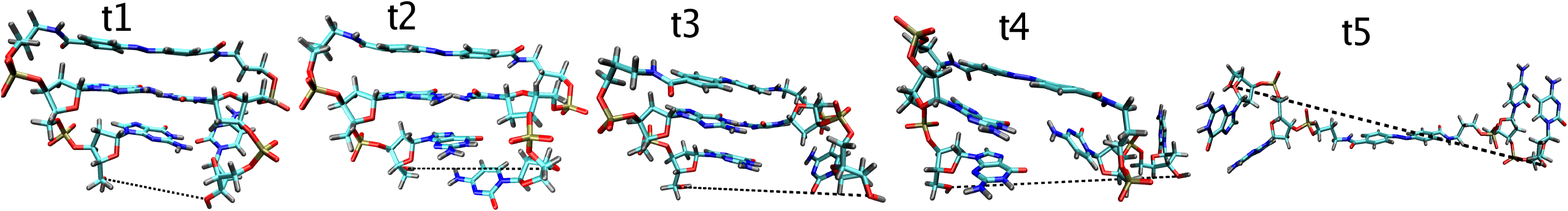}
   \includegraphics[width =0.65\textwidth]{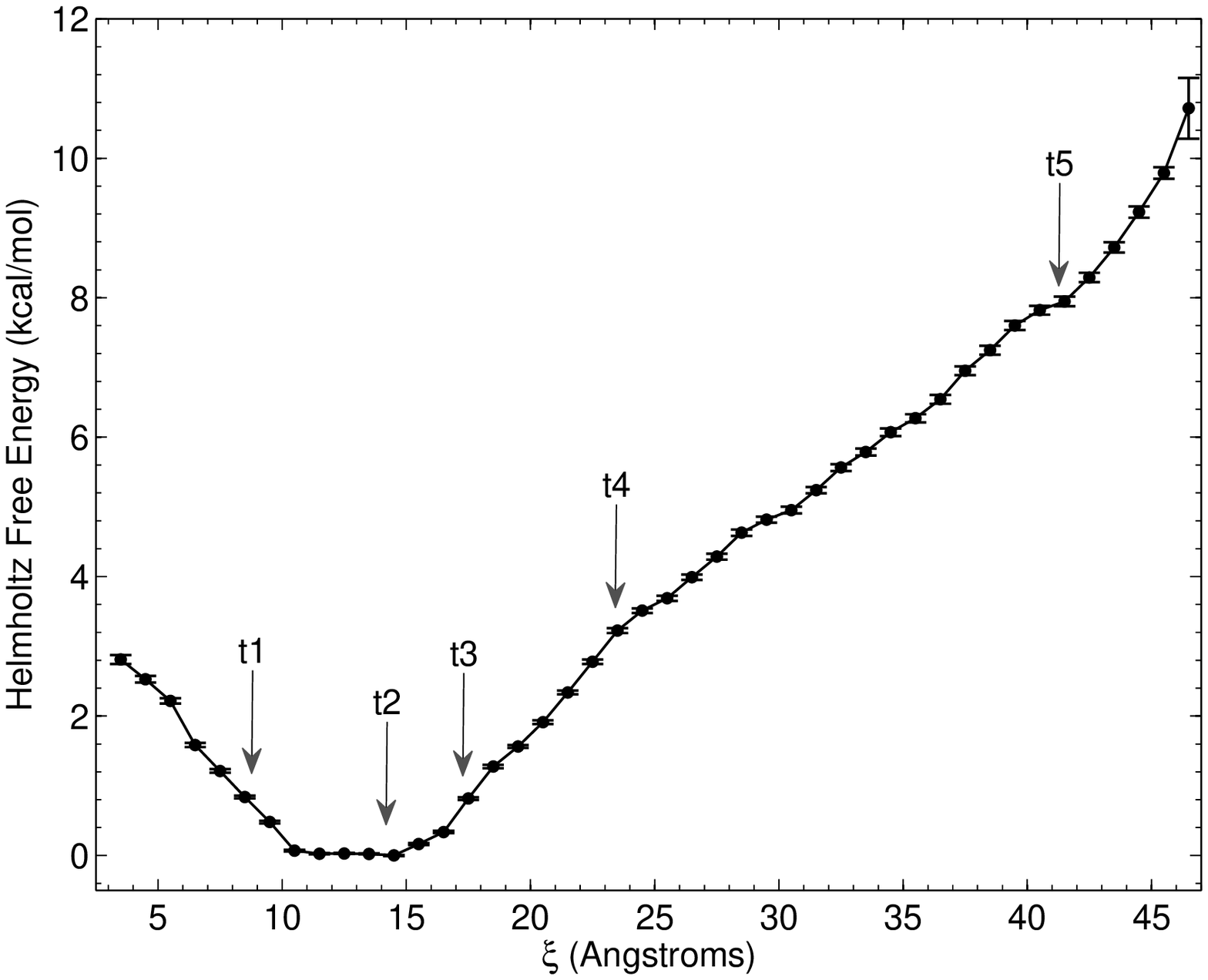}
   \caption{Potential of mean force along the end-to-end distance coordinate for the azobenzene capped DNA hairpin.  Snapshots of structures adopted during the unfolding are shown in the top panel. The error bars correspond to twice the standard deviation obtained from a bootstrapping analysis. }
   \label{fig:pmfdna}
   \end{figure}

Figure~\ref{fig:histogram} shows some of the  sampled  $p_L(\xi)$ [defined in \eq{eq:biased}]  that are used as input for the PMF reconstruction.  The resulting Helmholtz free energy profile  is shown in \fig{fig:pmfdna}. It consists of a convex region for $\xi<17$ \AA~ representing the folded state, a mostly concave region for 17 \AA$<
\xi <$ 40 \AA~  where the molecule unfolds, followed by another region of convexity that corresponds to the unfolded state. The unfolding region
exhibits some convex intervals signaling marginally stable intermediates encountered during the unfolding.  

\begin{figure}[htbp]
   \includegraphics[width =0.8\textwidth]{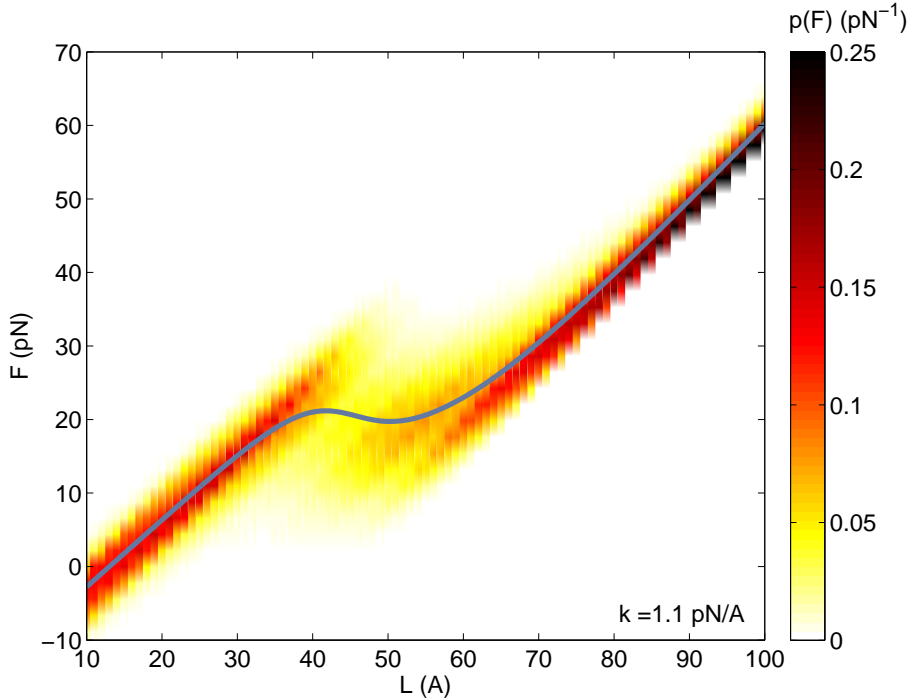}
   \caption{Force-extension profile of the DNA dimer when pulled with a cantilever of stiffness 1.1  pN/\AA~ estimated from the PMF in \fig{fig:pmfdna}. The solid line represents the average force-extension profile, the color plot the probability density distribution of the force measurements $p_L(F)$ [\eq{eq:Fprobdens}]. }
   \label{fig:forcedna}
   \end{figure}

Using the PMF one can then estimate the force-extension characteristics of the system when pulling with cantilevers of arbitrary stiffness as described in \stn{stn:forcefrompmf}. Figure~\ref{fig:forcedna} shows the predicted average and probability density in the force measurements  when the DNA dimer is pulled using a cantilever stiffness of 1.1 pN/\AA.  Note the degree of detail of the pulling process revealed by the probability density.

 %--------------------------------------------------%
 %--------------------------------------------------%
 
  %--------------------------------------------------%
 %--------------------------------------------------%

\section{Interpretation of pulling phenomenology}
\label{stn:interpretation}

A striking feature of single-molecule pulling is that the elastic behaviors of vastly different molecules often have qualitatively similar features.  In Secs.~\ref{stn:experiments} and~\ref{stn:modeling} we discussed two of these features: (i) mechanically unstable regions where $\frac{\partial \langle F\rangle_L}{\partial L} < 0$ and (ii) regions of dynamical bistability where blinking in the force measurements for fixed $L$ is observed.  In this section, we discuss the origin of both of these effects.  Emphasis  will be placed on the basic requirements on the molecule and the pulling device for the emergence of the phenomena. 

\subsection{Basic structure of the molecular potential of mean force}

\begin{figure}[htbp]
   \includegraphics[width =0.5\textwidth]{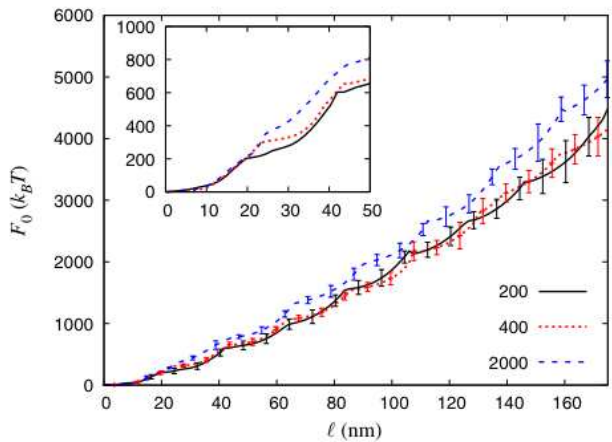}
   \caption{Molecular potential of mean force for a polyprotein composed of eight repeats of the Ig27 domain of human titin. The free energy profile was reconstructed using a nonequilibrium method and pulling at different pulling speeds of 200 (black curve), 400 (red dotted curve) and 2000 (blue dashed curve) nm/s. The number of nonequilibrium unfolding trajectories used in the estimate was 66, 35 and 29, respectively.  The force-extension profile of this molecule is shown in \fig{fig:imparato}. Figure taken from~\cite{imparato08}}
   \label{fig:imparatopmf}
   \end{figure}

Recall the basic structure of the PMF for the oligorotaxanes (\fig{fig:rotaxanePMF}): It consists of a convex region ($\frac{\partial^2 \phi(\xi)}{\partial \xi^2}> 0$) where the molecule is folded, another region of convexity when the molecule is extended and a region of concavity ($\frac{\partial^2 \phi(\xi)}{\partial \xi^2} < 0$)  where the molecule unfolds. This basic structure in the PMF is not unique to the rotaxanes but it is actually a common feature of the extension behavior of single molecules of sufficient structural complexity. Regions of convexity mark mechanically stable conformations while regions of concavity mark molecular unfolding events.  We had already encountered this very same basic structure when studying the elastic properties of the DNA dimer shown in \fig{fig:pmfdna}. Perhaps a more spectacular example of this structure is provided by the polyprotein composed of eight repeats of the Ig27 domain of human titin that we introduced in  \fig{fig:imparato}. The PMF reconstructed from experimental force-extension data is shown in \fig{fig:imparatopmf}. In this case also, mechanically stable conformations are described by regions of convexity in the PMF, while regions of concavity mark unfolding events. The reason why in this case there are 8 regions of concavity along the extension pathway is that there are 8 unfolding events, each marking the unfolding of one of the Ig27 domains in the polyprotein. Below we discuss the implications of these regions of concavity along the PMF. 

 %--------------------------------------------------%
 %--------------------------------------------------%

\subsection{Mechanical instability}

A first consequence of the regions of concavity along the molecular PMF is that they lead  to mechanically unstable regions in the $F$-$L$ isotherms. To see this  consider  the configurational partition function of the molecule plus cantilever at extension $L$ [\eq{eq:partfuncL2}] for large $k$. In this regime,  most contributions to the integral will come from the region where $\xi = L$. Consequently,   $\exp [ -\beta \phi(\xi) ]$ can be expanded around this point  to give:
 \be
\label{eq:stiffexpansion}
 \exp[-\beta \phi(\xi)]  = \exp [-\beta \phi(L) ] \left[  1 - \beta A_1(L) (\xi - L) 
 -\frac{\beta}{2} A_2(L)  (\xi -L)^2  + \cdots\right],
 \ee
 where $A_1(L)  =  \frac{\partial \phi(L)}{\partial L} $ and  
 $A_2(L)  =   \frac{\partial^2 \phi(L)}{\partial L^2} - \beta  \left( \frac{\partial \phi(L)}{\partial L} \right)^2$.
Here,  the notation is such that  
$ \frac{\partial \phi(L)}{\partial L} = \left.  \frac{\partial \phi(\xi)}{\partial \xi}\right|_{\xi = L}$. Introducing \eq{eq:stiffexpansion}  into \eq{eq:partfuncL2},  integrating explicitly the different terms,  and performing an expansion around $1/k = 0$ one obtains:
 \be
 Z(L) = \sqrt{\frac{2\pi}{k\beta}} \exp[ -\beta \phi(L) ]\left[ 1   -\frac{1}{2 k }  A_2(L)     + \mc{O}(1/k^2) \right].
 \ee
Using \eq{eq:averageforce}, the derivative of the force can be obtained from this approximation to the partition function. To zeroth order in  $1/k$ it is given by:
 \be
 \label{eq:stiffspring}
\frac{\partial \langle  F \rangle_L}{\partial L } = \frac{\partial^2 \phi (L)}{\partial L^2} + \mc{O}(1/k).
\ee
That is, an unstable region in the  force vs. extension where  $\frac{\partial \langle  F \rangle_L}{\partial L } <0 $ requires a region of concavity in the PMF. 

Note, however, that since the properties that are measured during pulling are those of the molecule plus cantilever, the observed mechanical stability properties will depend on the cantilever stiffness employed.  In fact, when employing very soft cantilevers no mechanically unstable region can be observed independent of the molecule that is being pulled, provided that no bond-breaking occurs during the pulling. 
To see this, consider the soft-spring approximation of the configurational partition function of the system plus cantilever [\eq{eq:partfuncL2}]:
\be
\label{eq:softspringZ}
\frac{Z(L)}{Z_0} \approx \frac{\exp( -\beta k L^2/2 )}{Z_0} \int \ud \xi \exp[ -\beta\phi(\xi) ] \exp( \beta k L \xi ) =  \exp( -\beta k L^2/2 )\langle  \exp( \beta k L \xi ) \rangle,
\ee
where the notation $\langle f \rangle$ stands for the unbiased (cantilever-free) average of $f$. In writing \eq{eq:softspringZ} we have supposed that in the region of relevant $\xi$ (in which the integrand is non-negligible)  $k\xi^2/V_L(\xi)\ll 1$ and, hence, that the cantilever potential is well approximated by $V_L(\xi) \approx \frac{k L^2}{2} - kL \xi$. This approximation is valid provided that that no bond breaking is induced during pulling and permits the introduction of a cumulant expansion~\cite{cumulant} in the configurational partition function. Specifically, the average $\langle  \exp( \beta k L \xi ) \rangle$ can be expressed as:
\be
\label{eq:cumulants}
 \langle \exp( \beta k L \xi ) \rangle = \sum_{n=0}^{\infty} \frac{(\beta k L)^n}{n!} \langle \xi^n\rangle  =  \exp \left\{ \sum_{n=1}^{\infty}  \frac{(\beta k L)^n}{n!}  \kappa_n (\xi) \right\} 
\ee
where $\kappa_n(\xi)$ is the $n$-th order cumulant. In view of  Eqs.~\eqref{eq:averageforce}, \eqref{eq:softspringZ} and  \eqref{eq:cumulants} the slope of the $F$-$L$ curves can be expressed as:
\be
\frac{\partial \langle F\rangle_L}{\partial L} = k - \beta k^2 \sum_{n=0}^{\infty} \frac{(\beta k L)^n}{n!} \kappa_{n+2}(\xi).
\ee
It then follows that to lowest order in $k$,
\be
\label{eq:decayinstability}
\frac{\partial \langle F \rangle_L}{\partial L} \approx k > 0.
\ee
That is, no unstable region in the $F$-$L$ curve can arise in the soft-spring limit, irrespective of the specific form of $\phi(\xi)$. 

\begin{figure}[htbp]
   \includegraphics[width =0.8\textwidth]{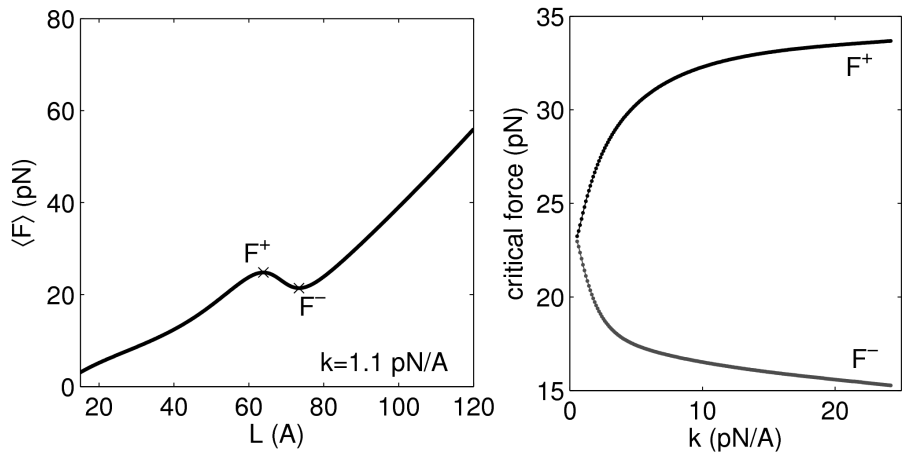}
   \caption{Dependence of the mechanical instability on the cantilever stiffness. Left panel: force-extension isotherm for a [3]rotaxane when pulled with a cantilever of stiffness $k=1.1$ pN/\AA. Here, $F^+$ and $F^-$ correspond to the values of the force when the $F$-$L$ curves exhibit a maximum and minimum,  and enclose the region of mechanical instability.  Right panel: dependence of $F^+$ and $F^-$  on $k$.  Note the decay of the region of mechanical instability for soft cantilevers. }
   \label{fig:instabilityvsk}
   \end{figure}

Figure~\ref{fig:instabilityvsk} illustrates the dependence of the region of mechanical instability on the cantilever stiffness using the [3]rotaxane case discussed in \stn{stn:oligorotaxane} as an example.  The figure shows the $k$-dependence of the local maximum $F^+$  and minimum $F^-$  in the average force measurements that enclose the region of mechanical instability.  The critical forces $F^+$ and $F^-$  show a  smooth and strong dependence on the cantilever spring constant. For large $k$  the system persistently  shows a  region of instability in the isotherms. However, as the cantilever spring is made softer, $F^+$ and $F^-$  approach each other and for small $k$ the mechanical instability in the $F$-$L$ isotherms is no longer present.

 %--------------------------------------------------%
 %--------------------------------------------------%
   
\begin{figure}[htbp]
\centering
\includegraphics[width=0.8\textwidth]{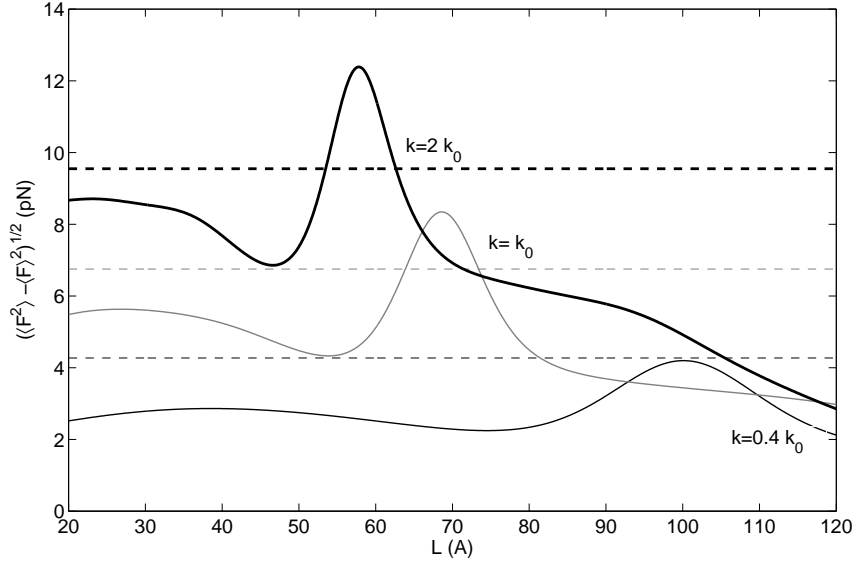}
\caption{Standard deviation  in the force measurements $\sigma_F = \sqrt{\langle F^2 \rangle_L - \langle F \rangle_L^2}$ as a function of $L$ for three different cantilever spring constants $k$ during the pulling of the [3]rotaxane. In each case, the dotted line indicates the value of $\sqrt{k/\beta}$ which sets the limit between the stable and unstable branches  in the extension, see \eq{eq:inequalities}. Here, $k_0=1.1$ pN/\AA. Figure taken from~\cite{franco09}. }
     \label{fig:fluctuations}
\end{figure}

Another interesting aspect of single-molecule pulling is that there is a close relationship between the mechanical stability properties and the fluctuations in the force measurements.  To see this,  note that  the  average slope of the $F$-$L$ curves can be expressed in terms of the  fluctuations in the force:
 \be
\label{eq:ffluctuations}
 \frac{\partial \langle F \rangle_L}{\partial L} = \left  \langle \frac{\partial^2 V_L(\xi)}{\partial L^2} \right \rangle_L
      - \beta [\langle F^2 \rangle_L - \langle F \rangle_L^2] =  k 
      - \beta [\langle F^2 \rangle_L - \langle F \rangle_L^2],
 \ee
where we have employed \eq{eq:averageforce}.
The sign of $ \frac{\partial \langle F \rangle_L}{\partial L}$   determines the mechanical stability during the extension and hence \eq{eq:ffluctuations}  relates the thermal fluctuations in the force measurements with the  stability properties of the $F$-$L$ curves. Specifically, for the stable regions for which $ \frac{\partial \langle F \rangle_L}{\partial L} >0$ the force fluctuations satisfy
\begin{subequations}
\label{eq:inequalities}
 \be
 \langle F^2 \rangle_L - \langle F \rangle_L^2 < \frac{k}{\beta}.
 \ee
In turn, in the unstable regions the force fluctuations are larger than in the stable regions and satisfy the inequality
\be
\label{eq:unstable}
 \langle F^2 \rangle_L - \langle F \rangle_L^2 > \frac{k}{\beta}.
\ee
\end{subequations}
The fact that the fluctuations are larger around the region of mechanical bistability is noticeable in some of the force-extension curves that we have discussed here  (see Figs.~\ref{fig:approachtoeql} and~\ref{fig:forcedna}). Figure~\ref{fig:fluctuations} illustrates these general observations in  the specific case of the pulling of [3]rotaxane. As shown, for  $k=k_0$ and $k=2k_0$ (where $k_0=1.1$ pN/\AA) there is a region where the force fluctuations become larger than $k/\beta$ and, consequently, unstable behavior in $F$-$L$ develops. For  $k=0.4k_0$ or less the fluctuations in the force are never large enough to  satisfy \eq{eq:unstable} and no critical points in the average $F$-$L$ curve develop, as can be confirmed in \fig{fig:instabilityvsk}. 

 %----------------------------------------------------------------%
 %--------------------------------------------------%

\subsection{Dynamical bistability}

\begin{figure}[htbp]
\centering
\includegraphics[width=0.8\textwidth]{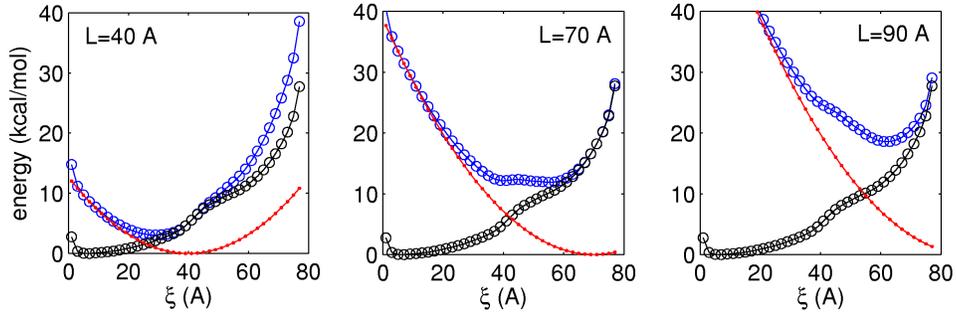}
\caption{Effective potential $U_L(\xi)=\phi(\xi) + V_L(\xi)$  for the molecule plus cantilever  for different values of the extension $L$. In the panels, the blue circles correspond to $U_L(\xi)$, the black circles to the PMF $\phi(\xi)$, and the solid line to the cantilever potential  $V_L(\xi)$ with $k=1.1$ pN/\AA.  Note the bistability in the effective potential for $L=70$ \AA.  } 
     \label{fig:emergencebistability}
\end{figure}

The observation of  force measurements that blink between a high-force and a low force regime for selected extensions requires the composite molecule plus cantilever system to be bistable along the end-to-end distance for some $L$. That is,  the effective potential $U_L(\xi)$ [\eq{eq:effectivepotential}] must have a double minimum.  Since the molecular PMF is not usually bistable along $\xi$ the bistability must be introduced by the cantilever potential.  To see how this bistability arises, consider the effective potential for the [3]rotaxane  for selected $L$ shown  in Figure~\ref{fig:emergencebistability}.  For $L$ in the mechanically stable regions of the force-extension isotherms  ($L=40$ \AA~ and $L=90$ \AA) the  effective potential exhibits a single minimum along the $\xi$ coordinate.  However, when $L=70$ \AA ~a bistability in the potential develops. At this extension the cantilever potential turns the region of concavity in $\phi(\xi)$ into a barrier between two minimum.  The secondary minimum  is the cause  for the blinking in the force measurements observed at this $L$ (cf. \fig{fig:3DNP2ring.unstable}). 

%----------------------------------------------------------------%
%----------------------------------------------------------------%

What are the minimum requirements for the emergence of bistability along $\xi$? A necessary condition  is that the effective potential $U_L(\xi)$ is concave for some region along $\xi$, that is:
\be
\label{eq:bistability}
\frac{\partial^2 U_L(\xi)}{\partial \xi^2} = \frac{\partial^2 \phi(\xi)}{\partial \xi^2} + k < 0.
 \ee
For \eq{eq:bistability} to be satisfied it is required that both: (i) the  PMF of the isolated molecule has  a region of concavity where $\frac{\partial^2 \phi(\xi)}{\partial \xi^2} < 0$, and (ii) the cantilever employed is sufficiently soft such that 
\be
\label{eq:bistabilityk}
k < -\ti{min} \left(\frac{\partial^2 \phi(\xi)}{\partial \xi^2}\right),
\ee
for some $\xi$.  Equation~\eqref{eq:bistabilityk} imposes an upper bound on $k>0$ for bistability to be observable. If $k$ is very stiff the  inequality would be violated for all $\xi$ and bistability would not be manifest. Note, however,  that there is no lower bound for $k$ that prevents bistability along $\xi$. 

\begin{figure}[htbp]
\centering
\includegraphics[width=0.8\textwidth]{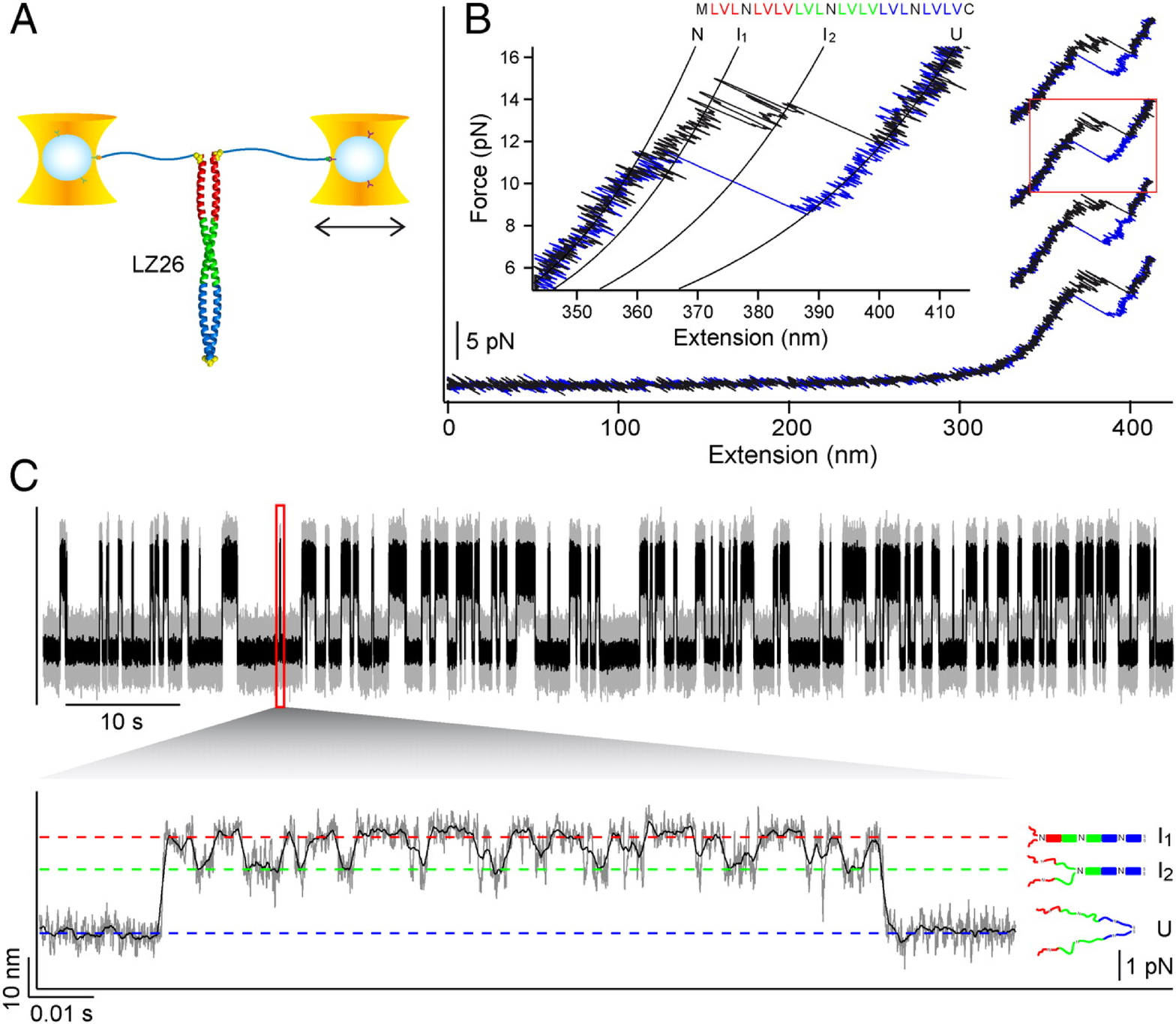}
\caption{Bistability during the pulling of a leucine zipper measured in an optical lattice arrangement. (A) Schematic of the experimental setup; (B) Sample force-extension traces; (C) Force vs time record for fixed $L$ in the region of dynamical bistability. Figure taken from~\cite{rief2010}.} 
     \label{fig:leucine}
\end{figure}

Since the region of concavity in the PMF is a common feature of molecules that have stable folded conformations, this bistability for selected extensions is a common feature of the pulling. We have already encountered this phenomenon during the pulling of the oligorotaxane and the  RNA hairpin (\fig{fig:rnapulling}). The DNA dimer (\fig{fig:forcedna}) and  the polyprotein in \fig{fig:imparato} are also expected to show this behavior since their PMF has regions of concavity along $\xi$.    The bistability has also been  predicted to be measurable when pulling $\pi$-stacked molecules~\cite{stacker, kim09}. For a discussion of the bistability in the isotensional ensemble, see Ref.~\cite{kirmizialtin2005}. Figure~\ref{fig:leucine} shows a striking recent experimental demonstration of the bistability obtained during the pulling of a leucine zipper~\cite{rief2010}.

%----------------------------------------------------------------%
%----------------------------------------------------------------%

\section{Summary}
\label{stn:conclusions}

The power of single-molecule pulling techniques is that they allow for mechanical control over the molecular conformation while simultaneously making thermodynamic measurements of any mechanically induced unfolding events.  The force-extension data often exhibits mechanically unstable regions where  the force decreases with increasing extension and bistable regions where, for fixed extension, blinks between a high-force and a low-force regime are observed. 

From the force-extension isotherms it is possible to reconstruct the PMF along the extension coordinate using the WHAM. In fact, the pulling process can be seen as an experimental realization of the WHAM methodology.  The PMF summarizes the changes in the free energy during folding  and determines the elastic properties of the molecule. Quite remarkably, the PMF's of widely different molecules often shares the same basic structure.  It consists of regions of convexity that represent mechanically stable molecular conformations interspersed by regions of concavity that signal molecular unfolding events.  

The starting point in the interpretation of single-molecule pulling experiments is \eq{eq:partfuncL2} which shows that the pulling process can be viewed as thermal motion along a one-dimensional potential $U_L(\xi)= \phi(\xi) + V_L(\xi)$ that is determined by the PMF $\phi(\xi)$ and the cantilever potential $V_L(\xi)$. Several important conclusions follow from this simple observation. The first and most obvious one is that the properties that are measured during pulling are those of the molecule plus cantilever  and hence that the basic phenomenon observed during pulling depend on the cantilever stiffness used.  It also follows that in order for the mechanical instability and dynamical bistability to be observable, the molecular PMF has to have a region of concavity. However, while the dynamical bistability survives for soft cantilever and decays for rigid cantilevers [see \eq{eq:bistabilityk}], the opposite is true for the mechanical instability [recall \eq{eq:decayinstability}].  The mechanical stability properties during pulling were also shown to be intimately related to the fluctuations in the force measurements [see \eq{eq:inequalities}]. 

One basic challenge in modeling the force spectroscopy using MD is to bridge the several orders of magnitude gap that exists between the pulling speeds used experimentally and those that are computationally feasible. We detailed two possible strategies that can be employed to overcome this difficulty: either strive for pulling speeds that are slow enough such that reversible behavior is recovered or use an indirect approach in which the PMF is first reconstructed and the force-extension behavior is then estimated from the PMF. Other strategies based on nonequilibrium pulling are also possible.  From the simulation of the pulling one can gain insights into the conformations encountered during molecular unfolding and into structure-function relations responsible for the elastic behavior of single molecules.

Currently, the accurate reconstruction of the PMF of molecular systems using equilibrium sampling is a serious computational challenge except for relatively modest systems.  The recent discovery of nonequilibrium work fluctuation relations has led to a surge of activity in seeking nonequilibrium methods to reconstruct the PMF that can be used in situations where proper equilibrium sampling is unfeasible. The full consequences of this venue of research is still to be determined.

\begin{acknowledgments}
This work was supported by the Non-equilibrium Energy  Research Center (NERC) which is an Energy Frontier Research Center funded by the U.S.
Department of Energy, Office of Science, Office of Basic Energy Sciences under Award Number DE-SC0000989. The authors thank Dr. Martin McCullagh for useful discussions.  
\end{acknowledgments}

\bibliography{francobook}
\end{document}